\documentclass[final,3p,times,twocolumn]{elsarticle}

\usepackage{graphicx}
\usepackage{dcolumn}
\usepackage{amssymb}
\usepackage{amsthm}
\usepackage[dvips]{pstcol}
\usepackage{mathtools}

\usepackage{fancybox}

\newcounter{bla}

\newcommand{\mnotex}[1]
{\protect{\stepcounter{mnotecount}}$^{\mbox{\footnotesize$
\bullet$\themnotecount}}$ \marginpar{
\raggedright\tiny\em
$\!\!\!\!\!\!\,\bullet$\themnotecount: #1} }

\newcounter{mnotecount}

\newcommand{\DAl}{\raise -0.4mm \hbox{\large$\Box$}}

\newcommand{\dd}{\;\rule{0.018cm}{.15cm}\!\raisebox{1mm}{\rule{0.08cm}{.02cm}}}

\newcommand{\ddn}{\;\rule{0.018cm}{.25cm}\!\raisebox{1.7mm}{\rule{0.1cm}{.02cm}}}

\newcommand{\mathsym}[1]{{}}
\newcommand{\unicode}[1]{{}}

\newcommand{\icode}[1]{{\scriptsize\tt #1}}

\newcounter{mathex}
\newcommand{\Mathsize}{\scriptsize}
\newcommand{\MathIn}[2]{ {\textcolor{blue}{\tt In[\arabic{#1}]:=} }\ { \tt #2 } \\[1mm] }
\newcommand{\MathOut}[1]{{\textcolor{blue}{\tt Out[\arabic{#1}]=}}\ \ }

\newcommand{\DefTensorMessage}[1]{ {\tt ** DefTensor: Defining tensor #1.} \\ }

\newcommand{\link}{$\underbracket[.2mm][.4mm]{\ \ }_{\ }$}

\newtheorem{theorem}{Theorem}

\newtheorem{definition}{Definition}

\journal{Computer Physics Communications}

\begin{document}

\begin{frontmatter}



\title{{\em Spinors}: a Mathematica package for doing spinor calculus in General Relativity}


\author[a]{Alfonso Garc\'{\i}a-Parrado G\'omez-Lobo\corref{author}}
\author[b,c]{Jos\'e M. Mart\'{\i}n-Garc\'{\i}a\fnref{WRI}}

\cortext[author] {Corresponding author.\\\textit{E-mail address:} alfonso@math.uminho.pt}

\address[a]{Centro de Matem\'atica, Universidade do Minho. 4710-057 Braga, Portugal}
\address[b]{Laboratoire Univers et Th\'eories, Observatoire de Paris, CNRS,
Univ. Paris Diderot, 5 place Jules Janssen, 92190 Meudon, France}
\address[c]{Institut d'Astrophysique de Paris, Univ. Pierre et Marie Curie, CNRS,
98bis boulevard Arago, 75014 Paris, France
}

\fntext[WRI]{Current address: Wolfram Research Inc., 100 Trade Center Dr., Champaign IL 61820, USA}




\begin{abstract}
The {\em Spinors} software is a {\em Mathematica} package which implements 
2-component spinor calculus as devised by Penrose for General Relativity in 
dimension 3+1. 
The {\em Spinors} software is part of the {\em xAct} system, which is a collection of {\em Mathematica} packages to do tensor analysis by computer. 
In this paper we give a thorough 
description of {\em Spinors} and present practical examples of use.

\end{abstract}


\end{frontmatter}



{\bf PROGRAM SUMMARY/NEW VERSION PROGRAM SUMMARY}

\begin{small}
\noindent
{\em Manuscript Title:} Spinors: a Mathematica package for doing spinor calculus in General Relativity. \\
{\em Authors:} A. Garc\'{\i}a-Parrado and J.M. Mart\'{\i}n-Garc\'{\i}a. \\
{\em Program Title:} Spinors.                                  \\
{\em Journal Reference:}                                      \\
{\em Catalogue identifier:}                                   \\
{\em Licensing provisions:}                                   \\
{\em Programming language:} Mathematica.                        \\
{\em Computer:} Any computer running Mathematica 7.0 or higher. \\
{\em Operating system:} Any operating system compatible with Mathematica 7.0 or higher. \\
{\em RAM:} 94Mb in Mathematica 8.0.                                              \\
{\em Number of processors used:} 1.                             
{\em Classification:} Relativity and Gravitation. \\
{\em External routines/libraries:} Mathematica packages xCore, xPerm and xTensor which are part 
of the xAct system. These can be obtained at http://www.xact.es.\\
{\em Nature of problem:} Manipulation and simplification of spinor expressions in 
General Relativity.\\
   \\
{\em Solution method:} Adaptation of the tensor functionality of the xAct system for the specific situation of spinor calculus in four dimensional Lorentzian geometry.\\
   \\
{\em Restrictions:} The software only works on 4-dimensional Lorentzian space-times with 
metric of signature $(1,-1,-1,-1)$. There is no direct support for Dirac spinors.\\
   \\
{\em Unusual features:} Easy rules to transform tensor expressions into spinor ones and back.
Seamless integration of abstract index manipulation of spinor expressions with component computations. \\
   \\
 {\em Running time:} Under one second to handle and canonicalize standard spinorial expressions with a few dozen indices. (These expressions arise naturally in the transformation of a spinor expression into a tensor one or vice-versa).\\
   \\
* Items marked with an asterisk are only required for new versions
of programs previously published in the CPC Program Library.\\
\end{small}

\section{Introduction} 
The concept of {\em spinor} plays an important role in certain areas of mathematical and 
theoretical physics. Roughly speaking a spinor is a field which transforms under a 
{\em spinor representation} of a given symmetry group in our system. 
For example, if we are working in a pseudo-Riemannian manifold with a metric of 
signature $(p,q)$ ($p$ represents the number of $+1$ and $q$ the number of 
$-1$ entries in the canonical form of the metric),
then a natural symmetry group is the group which transforms orthonormal 
frames into orthonormal frames. This group is $O(p,q)$ ($SO(p,q)$ if we restrict ourselves to transformations preserving the 
{\em frame orientation}). The spin group 
is then the universal covering of
$SO(p,q)$ which, as is well-known, is $\mbox{\em Spin}(p,q)$ and hence 
spinors transform under irreducible representations of this group.

The above considerations are completely general and they enable us to introduce
the notion of spinor field in any pseudo-Riemannian manifold admitting a 
{\em spin structure}. However, in the case of a 4-dimensional Lorentzian 
manifold (the space-time model in
General Relativity) a more algebraic approach is desired. This approach was pioneered 
by Penrose \cite{PENROSE-SPINORS} where he stu\-di\-ed the main properties of the spinor
algebra of those spinors arising from the spin group of $SO(1,3)$ and in addition  
he developed a calculus adapted to the par\-ti\-cular spin vector bundle which one can define in a
4-dimensional Lorentzian manifold admitting a spin structure. Penrose's spinor calculus revealed very useful
in certain contexts of General Relativity (GR) where the use of tensor 
methods results in very cumbersome computations. 
Perhaps the best known example is the spinor formulation of the algebraic classification of the 
Weyl tensor (Petrov classification). The spinor form of the Weyl tensor is a totally symmetric 
4-rank spinor and it is very easy to show that such a spinor can only admit four different 
algebraic types which are in correspondence with the four distinct Petrov types. 

In this article we describe the {\em Mathematica} package {\em Spinors} which 
implements the spinor calculus in four dimensional Lorentzian geometry as
conceived by Penrose. In this conception, spinors are tensor fields
on a certain tensor bundle and therefore one can use the general ideas
of tensor bundles to work with spinors. In particular the notion of 
spin covariant derivative, the curvature spinors or the relation between spinors
and space-time tensors find here a natural formulation. An important part of 
this formulation is the notion of {\em abstract index} used to represent tensor
fields on any tensor bundle. This representation of tensor fields  has been 
adopted in the system {\em xAct} \cite{XACT}, which {\em Spinors} is part of. 
The system {\em xAct} is a system to do tensor analysis by computer in
{\em Mathematica}, both by working with tensors as linear combination
of basis tensors (component calculus) and by working with tensors as
symbolic names with certain properties like rank or symmetry (abstract
calculus). The system {\em xAct} consists of different modules tailored
for different tasks and {\em Spinors} is one of these modules.     

Other computer algebra systems support computations with spinors.
For example in the context of Particle Physics we may quote the package
{\em Spinors@Mathematica} \cite{MS} which can be used in the evaluation
of scattering amplitudes at tree and loop level.
The stand-alone package {\em Cadabra} \cite{CADABRA} handles generic
abstract spinor quantities in any dimension, with emphasis in Field
Theory, but no special support for General Relativity or component
computations.
The Maple built-in package {\em DifferentialGeometry} has extensive
support for component computations of multiple types, in particular
the NP formalism, but no support for abstract tensor computations.
Another Maple package handling the NP formalism is {\em NPSpinor}
\cite{NPSPINOR}.

The paper is organised as follows: in section \ref{sec:preliminaries} 
we give a 
mathematical introduction to spinor calculus. The aim of this introduction 
is to set the notation and conven\-tions which are followed by the {\em Spinors} 
implementation. Section \ref{sec:xact} explains how the {\em Spinors} software
fits into the {\em xAct} framework and section \ref{sec:spinors-description} presents
a practical session with {\em Spinors} in which the main features of 
the program are shown by means of practical examples. The paper is finished in 
section \ref{sec:sparling} where a practical computation involving the 
{\em Nester-Witten} spinor and the {\em Sparling identity}  
is carried out with
{\em Spinors}.

\section{Mathematical preliminaries}
\label{sec:preliminaries}
In this section we give an overview of the spinor calculus in General Relativity, 
following a practical approach to introduce the subject and omitting most of the proofs
(detailed studies can be found in e.g. \cite{PR-RINDLER-1,ASHTEKAR}). 
Let $\mathbf L$ be a 4-dimensional real vector space endowed with a real scalar product 
$g(\ ,\ )$ of Lorentzian signature and let $\mathbf S$ be a 2-dimensional complex vector space (complex conjugate of scalars
will be denoted by an overbar). 
The vector space $\mathbf S$ is related to another complex vector space $\overline{\mathbf S}$
by an anti-linear, involutive transformation.  

The vector space $\mathbf L$ and its dual ${\mathbf L}^*$ can be used as the starting point to 
build a tensor algebra in the standard fashion. Similarly a tensor algebra is built 
from $\mathbf S$, $\overline{\mathbf S}$ and their respective duals ${\mathbf S}^*$, 
$\overline{\mathbf S}^*$. We denote these algebras by $\mathfrak{T}(\mathbf{L})$, 
$\mathfrak{T}(\mathbf{S})$ and 
$\mathfrak{T}(\overline{\mathbf{S}})$ respectively 
\footnote{Strictly speaking only the algebras $\mathfrak{T}^r_s(\mathbf L)$ 
of tensors $r$-contravariant $s$-covariant can be defined (and the same applies 
to $\mathfrak{T}^{r}_{s}(\mathbf S)$). To lessen the notation we will suppress the 
labels $r$, $s$ in the notation and they will only be made explicit when confusion may arise.}. 
In this work abstract indices will be used throughout to denote tensorial quantities: in this 
way lowercase Latin indices $a,b,\dots$ will denote abstract indices on elements of 
$\mathfrak{T}(\mathbf{L})$ and capital Latin indices $A,B,\dots$ 
(resp. primed capital Latin indices $A',B',\dots$) will be used for 
abstract indices of elements in $\mathfrak{T}(\mathbf{S})$ 
(resp. $\mathfrak{T}(\overline{\mathbf{S}})$). 
The union of the tensor algebras $\mathfrak{T}(\mathbf{S})$, 
$\mathfrak{T}(\overline{\mathbf{S}})$ will be referred to as the 
{\em spin algebra} and its elements will be called spinors. 
One can also build tensor algebras by taking tensor products of 
elements in $\mathfrak{T}(\mathbf L)$, $\mathfrak{T}(\mathbf S)$ 
and $\mathfrak{T}(\overline{\mathbf{S}})$. Quantities in these 
tensor algebras will be referred to as {\em mixed quantities} 
and they will carry abstract indices of tensor and spinor type. 
All tensor algebras shall be regarded as complex vector spaces.

Since ${\mathbf S}$ is 2-dimensional, we deduce that the vector space of 
antisymmetric 2-spinors is 1-dimensional and therefore we can pick up a 
non-vanishing representative $\epsilon_{AB}$ which generates such a vector space. 
We define next a spinor $\widehat{\epsilon}^{AB}$ by the relation
\begin{equation}
\epsilon_{AB}\widehat\epsilon^{CB}=\delta^{\phantom{A}C}_{A}.
\label{eq:epsinverse}
\end{equation}
where $\delta_{A}^{\phantom{A}C}$ is the identity tensor (also known as the 
{\em Kronecker delta}) on the vector space $\mathbf{S}$. Indeed the spinors $\epsilon_{AB}$, 
$\widehat{\epsilon}^{AB}$ can be used to relate elements in $\mathbf{S}$ and elements in
$\mathbf{S}^*$ in the following way
\begin{equation}
\xi^A\epsilon_{AB}=\xi_B\;,\quad \xi^A=\widehat\epsilon^{AB}\xi_B\;,
\label{eq:raise-lower}
\end{equation}
where $\xi^A$ is an arbitrary spinor in $\mathbf{S}$.     
Hence, the spinors $\epsilon_{AB}$ and $\widehat{\epsilon}^{AB}$ can be understood as 
a metric on ${\mathbf S}$ (symplectic metric) and its inverse and the operation shown in (\ref{eq:raise-lower}) is the standard 
``raising and lowering'' of indices.
These operations are extended to the full spinor algebra without difficulty.
In particular we can raise the indices of $\epsilon_{AB}$ getting 
$\epsilon^{AB}=\widehat\epsilon^{AB}$ and from now on only the symbol $\epsilon$ will be used for the symplectic metric and its inverse. Note also the property 
\begin{equation}
\delta^A_{\phantom{A}B}=-\delta_B^{\phantom{B}A}.
\label{eq:kronecker-delta}
\end{equation}
Here the quantity $\delta_B^{\phantom{B}A}$ is the Kronecker delta on 
$\mathbf S$ and $\delta^A_{\phantom{A}B}$ is a derived quantity obtained from it by the raising and lowering of indices. In particular this implies $\delta_A^{\phantom{A}A}=2$. The spinors $\epsilon_{AB}$, $\epsilon^{AB}$ and $\delta_{A}^{\phantom{A}B}$ all have counterparts 
(complex conjugates) defined in the algebra $\mathfrak{T}(\overline{\mathbf{S}})$.

It is possible to relate tensors and spinors by means of the {\em soldering form}. This is a mixed quantity  $\sigma_a^{\phantom{a}AA'}$ fulfilling the algebraic properties
\begin{eqnarray*}
&&\sigma^a_{\phantom{a}AA'}\sigma_{aBB'}=\epsilon_{BA}\bar{\epsilon}_{B'A'}
\;,\quad \sigma^a_{\ AA'}\sigma_b^{AA'}=\delta^a_{\ b}\;,\\
&&\overline{\sigma_a^{\phantom{a}AA'}}=\sigma_a^{\phantom{a}AA'}.
\end{eqnarray*}
The last of these properties implies that $\sigma^a_{\phantom{a}AA'}$ is 
{\em hermitian}. This is only compatible with the metric signature $(1,-1,-1,-1)$. Choosing $\sigma^a_{\phantom{a}AA'}$ anti-hermitian would be only compatible with the signature $(-1,1,1,1)$ \cite{ASHTEKAR}. 
These properties enable us to relate tensors and spinors in the following way

\begin{eqnarray}
&&\hspace{-1.5cm}T^{A_1A_1'\dots A_pA'_p}_{\ B_1B'_1\dots B_qB_q}
=T^{a_1\dots a_p}_{\ b_1\dots b_q}
\sigma^{b_1}_{\phantom{b_1}B_1B'_1}\cdots\sigma^{b_q}_{\phantom{b_q}B_qB'_q}
\sigma_{a_1}^{\phantom{a_1}A_1A'_1}\cdots\sigma_{a_p}^{\phantom{a_p}A_pA'_p},\nonumber\\ 
&&\label{ivv}\\
&&\hspace{-1.5cm}T^{a_1\dots a_p}_{\  b_1\dots b_q}=T^{ A_1A_1'\dots A_pA'_p}_{\  B_1B'_1\dots B_qB_q}
\sigma_{b_1}^{\phantom{b_1}B_1B'_1}\cdots\sigma_{b_q}^{\phantom{b_q}B_qB'_q}
\sigma^{a_1}_{\phantom{a_1}A_1A'_1}\cdots\sigma^{a_p}_{\phantom{a_p}A_pA'_p}.
\nonumber\\
&&\label{vvi}\;
\end{eqnarray}
 where $T^{a_1\dots a_p}_{\  b_1\dots b_q}$ is an arbitrary tensor and 
$T^{A_1A_1'\dots A_pA'_p}_{\ B_1B'_1\dots B_qB_q}$ its spinor counterpart 

Another important algebraic property of the soldering form is
\begin{equation}
\sigma {}_{a}{}^{A}{}_{A'} \sigma {}_{b}{}_{A}{}_{C'} + \sigma \
{}_{a}{}_{A}{}_{C'} \sigma {}_{b}{}^{A}{}_{A'} =  g{}_{a}{}_{b} \
\overline{\epsilon }{}_{A'}{}_{C'}.
\label{eq:clifford}
\end{equation}
This equation is a direct consequence of the irreducible decomposition of the product 
$\sigma{}_{a}{}^{A}{}_{A'}\sigma {}_{b}{}_{A}{}_{C'}$ according to theorem
\ref{theo:irr-decomp} below and the algebraic properties of the soldering form.
Starting from (\ref{eq:clifford}) we can derive formulas for the products of soldering forms 
with all their spinor indices contracted 
(these are useful to translate spinor expressions into tensor ones). For example
\begin{eqnarray}
&&\sigma {}_{a}{}_{A}{}_{A'} \sigma {}_{b}{}^{A}{}_{C'} 
\sigma{}_{c}{}^{D}{}^{C'} \sigma {}_{d}{}_{D}{}^{A'}=\nonumber\\
&&\frac{1}{2}(\mbox{i}\eta{}_{c}{}_{d}{}_{a}{}_{b} + g{}_{a}{}_{b} g{}_{c}{}_{d} + g{}_{c}{}_{b} g{}_{d}{}_{a}- g{}_{c}{}_{a} g{}_{d}{}_{b})\;,
\label{eq:sigma-product}
\end{eqnarray}
where $\eta_{abcd}$ is the volume form of the metric $g_{ab}$. 
It is possible to generalise this formula for the case  of a product of more soldering forms. These can be written as contracted products of the quantity
\begin{equation}
G_{abcd}\equiv \frac{1}{2}
(\mbox{i}\eta{}_{a}{}_{b}{}_{c}{}_{d} + g{}_{a}{}_{d} g{}_{b}{}_{c} 
- g{}_{a}{}_{c} g{}_{b}{}_{d}+ g{}_{a}{}_{b} g{}_{c}{}_{d}).
\label{eq:tetra-g}
\end{equation}
Combining eq. (\ref{eq:sigma-product}) and its complex conjugate we obtain the spinor counterpart of $\eta_{abcd}$, written as follows
\begin{eqnarray}
&& \eta{}_{a}{}_{b}{}_{c}{}_{d} \sigma {}^{a}{}_{A}{}_{A'} 
\sigma{}^{b}{}_{B}{}_{B'} \sigma {}^{c}{}_{C}{}_{C'}\sigma{}^{d}{}_{D}{}_{D'} =\nonumber\\
&&i (\epsilon {}_{A}{}_{C}\epsilon{}_{B}{}_{D} \overline{\epsilon }{}_{A'}{}_{D'} \
\overline{\epsilon }{}_{B'}{}_{C'} -\epsilon{}_{A}{}_{D}\epsilon{}_{B}{}_{C} \overline{\epsilon }{}_{A'}{}_{C'}
\overline{\epsilon }{}_{B'}{}_{D'}). 
\end{eqnarray}
We finish this review about spinor algebra by recalling an important result dealing with the 
decomposition of an arbitrary spinor into irreducible parts 
under the Lorentz group \cite{PR-RINDLER-1}. 

\begin{theorem}
 Any spinor $\xi_{A_1\cdots A_p B'_1\cdots B'_q}$, $p,q\in\mathbb{N}$ can be written as the sum
of a totally symmetric spinor $\xi_{(A_1\cdots A_p) (B'_1\cdots B'_q)}$ plus terms which are 
products of the spin metric $\epsilon_{AB}$ (or its complex conjugate $\bar\epsilon_{A'B'}$) 
times totally symmetric spinors of lower rank.
\label{theo:irr-decomp}
\end{theorem}

\subsection{Spinor calculus}
\label{subsec:spinor-calculus}
So far all our considerations were algebraic in nature, but we can also perform our construction for the case of a Lorentzian manifold $(\mathcal{M}, g)$ as follows: the construction performed in previous paragraphs is carried out taking as vector space ${\mathbf L}$ the tangent space $T_p(\mathcal{M})$ of an arbitrary point $p\in\mathcal{M}$ which is endowed with the Lorentzian scalar product $g|_p$. In this way it is possible to introduce a complex vector space $\mathbf{S}_p$ and a quantity $\sigma_a^{\phantom{a}AA'}|_p$. Now the set 
$S(\mathcal{M})\equiv\bigcup_{p\in{\mathcal M}} {\mathbf S}_p$ is a vector bundle with the manifold $\mathcal{M}$ as the base space and the group
of linear transformations on $\mathbb C^2$ as the structure group. We will call this vector bundle the {\em spin bundle} and the sections of $S(\mathcal{M})$ are the contravariant rank-1 spinor fields on ${\mathcal M}$.
We can now define the tensor algebras $\mathfrak{T}^r_s(T_p({\mathcal M}))$, 
$\mathfrak{T}^R_S(\mathbf S_p)$ and, 
use them to construct vector bundles with $\mathcal M$ as the base manifold. 
These bundles are tensor bundles and we denote each of these tensor bundles by 
$\mathfrak{S}^{r,R}_{s,S}(\mathcal M)$, where the meaning of the labels $r$, $R$, $s$, $S$ is the obvious one.  In general we will suppress these labels and use just the notation 
$\mathfrak{S}(\mathcal M)$ as a generic symbol for these tensor bundles. Sections on $\mathfrak{S}(\mathcal M)$ are written using abstract indices and we follow the same conventions as in the case of the vector spaces ${\mathbf L}$ and 
${\mathbf S}$. Sections of any of the bundles $\mathfrak{S}^{0,R}_{0,S}(\mathcal M)$
are called {\em spinor fields} or simply {\em spinors}. As usual there is a complex conjugate counterpart of this bundle, denoted by $\overline{\mathfrak{S}(\mathcal M)}$.
\begin{definition}[Spin structure]
If the quantity  $\sigma_{a}^{\phantom{a}AA'}|_p$ varies smoothly on the manifold $\mathcal M$, then one can define a smooth section, denoted by $\sigma_{a}^{\phantom{a}AA'}$. 
When this is the case we call the smooth section $\sigma_{a}^{\phantom{a}AA'}$
a smooth spin structure on the Lorentzian manifold $({\mathcal M},g)$. 
\end{definition}
Clearly a spin structure can be always defined in a neighbourhood of any point $p\in\mathcal{M}$ but further topological restrictions are required if the spin structure is to be defined globally (see e.g. \cite{NAKAHARA}). 

We turn now to the study of covariant derivatives defined on the bundles 
$\mathfrak{S}(\mathcal M)$, $\overline{\mathfrak{S}(\mathcal M)}$. Let $D_a$ denote such a covariant derivative. Then the operator $D_a$ can act on any quantity with tensor indices and/or spinor indices. As a result, when $D_a$ is restricted to quantities having only tensor indices  we recover the standard notion of covariant derivative acting on tensor fields of $\mathcal M$. If $D_a$ is restricted to quantities having only spinor indices then $D_a$ is the covariant derivative acting on spinor fields. The consequence of this is that the connection coefficients and the curvature of $D_a$ will be divided in two groups: quantities arising from the tensorial part and quantities arising from the spinorial part. The group arising from the tensorial part consists of the Christoffel symbols/Ricci rotation coefficients and the Riemann tensor of the covariant derivative restricted to the tangent bundle $T(\mathcal M)$. The group coming from the spinorial part contains the connection components and the curvature tensor of the covariant derivative restricted to the spin bundle $S(\mathcal M)$ (or $\overline{S(\mathcal M)}$). 
We will refer to these as the {\em inner connection} and the {\em inner curvature} respectively. See \cite{ASHTEKAR,ASHTEKAR-CALCULUS} for an in-depth discussion of these concepts.

\begin{definition}[\bf Spin covariant derivative]
Suppose that $\mathfrak{S}(\mathcal M)$ admits a spin structure 
$\sigma_{b}^{\phantom{b}AA'}$. We say that a covariant derivative $D_a$ defined on 
$\mathfrak{S}(\mathcal M)$ is compatible with the spin structure $\sigma_{b}^{\phantom{b}AA'}$ 
if it fulfils the property  
\begin{equation}
D_a\sigma_{b}^{\phantom{a}AA'}=0.
\label{eq:spin-structure-condition}
\end{equation}
The covariant derivative $D_a$ is then called a spin covariant derivative with respect to the spin structure $\sigma_{b}^{\phantom{a}AA'}$.
\label{def:spin-covd}
\end{definition}
Given that any quantity antisymmetric in two spinor indices must contain the spin metric as 
a factor we have 
\begin{equation}
D_a\epsilon_{AB}=\lambda_a\epsilon_{AB}\;,\quad 
\lambda_a\equiv\epsilon^{AB}D_a\epsilon_{AB}\;,
\label{eq:covd-epsilon}
\end{equation}
where $D_a$ is any covariant derivative defined on the bundle $\mathfrak{S}(\mathcal M)$.
When $D_a$ is in addition a spin covariant derivative then, combining (\ref{eq:spin-structure-condition}) and (\ref{eq:covd-epsilon}) we easily deduce the additional properties 
\begin{equation}
D_a\sigma^b_{\phantom{b}AA'}=0\;,\quad D_c g_{ab}=(\lambda_c+\bar\lambda_c)g_{ab}. 
\end{equation}
The last equation implies that any spin convariant derivative gives rise to a 
{\em semi-metric} connection \cite{SCHOUTEN} when restricted to the space-time tensor bundle. 
If furthermore $D_a$ has no torsion, then it is known as a {\em Weyl connection}.
The spin covariant derivative  
gets fixed if we demand additional properties on it (see e. g. \cite{PR-RINDLER-1}) 
\begin{theorem}
There is one and only one torsion-free spin covariant derivative 
$\nabla_a$ with respect to the spin structure 
$\sigma_{a}^{\phantom{a}AA'}$ which fulfills the property
\begin{equation}
\nabla_a\epsilon_{AB}=0.
\label{eq:cd-epsilon}
\end{equation}
\label{theorem:spincd-unique}
\end{theorem}
Acting with such $\nabla_a$ on (\ref{eq:clifford}) gives
\begin{equation}
\nabla_a g_{bc} = 0,
\label{eq:SpinLeviCivita}
\end{equation}
which shows that the restriction of $\nabla_a$ to quantities with tensorial
indices is just the Levi-Civita covariant derivative of $g_{ab}$.

Consider now any spinor field $\xi^A$ and any spin covariant derivative $D_a$. 
Then the commutation of $D_a$, $D_b$ acting on $\xi^A$ is given by \cite{ASHTEKAR}
\begin{equation}
D_aD_b\xi^B-D_bD_a\xi^B-T^r_{\phantom{r}ab}D_r\xi^B=F_{abR}^{\phantom{abC}B}\xi^C
\label{eq:ricci-identity-F}
\end{equation}
where $T^r_{\phantom{r}ab}$ is the torsion of $D_a$. The mixed quantity $F_{abA}^{\phantom{abA}B}$ is the inner curvature mentioned above. It is antisymmetric in the tensorial indices and it fulfils the {\em Bianchi identity} \cite{ASHTEKAR}
\begin{equation}
D_{[a}F_{bc]\phantom{B}A}^{\phantom{bc]}B}+T_{\phantom{r}[ab}^rF_{c]r\phantom{B}A}^{\phantom{c]r}B}=0.
\label{eq:F-bianchi-identity} 
\end{equation}
The spinor counterpart of the inner curvature is represented by 
$F_{CC'DD'A}^{\phantom{CC'DD'A}B}$ and it can be decomposed as follows
\begin{equation}
F_{CC'DD'AB}=X_{ABCD}\bar{\epsilon}_{C'D'}+\Phi_{ABC'D'}\epsilon_{CD}. 
\label{curvature-spinors} 
\end{equation}
The spinors $X_{ABCD}$ and $\Phi_{ABC'D'}$ are called {\em curvature spinors} and they enjoy the symmetries
\begin{equation}
X_{AB(CD)}=X_{ABCD}\;,\quad
\Phi_{AB(C'D')}=\Phi_{ABC'D'}. 
\end{equation}
We can also introduce a spinor $T^{AA'}_{\phantom{AA'}BB'CC'}$ representing the torsion. Its irreducible decomposition reads
\begin{equation}
T^{AA'}_{\phantom{AA'}BB'CC'}=\Omega^{AA'}_{\phantom{AA'}BC}\bar{\epsilon}_{B'C'}
+\bar{\Omega}^{AA'}_{\phantom{AA'}B'C'}\epsilon_{BC}, 
\label{eq:torsion-decomposition}
\end{equation}
where $\Omega_{AA'BC}$ is the {\em torsion spinor} and it fulfills the symmetries
\begin{equation}
\Omega_{AA'(BC)}=\Omega_{AA'BC}.
\end{equation}
The inner curvature and the Riemann tensor of $D_a$ are indeed related. To find the relation between them one computes the Ricci identity for an arbitrary vector $V^a$ and then particularises it for the special case in which $V^a=\sigma^a_{\phantom{a}AA'}\xi^A\xi^{A'}$.
The result is 
\begin{equation}
R_{abc}^{\phantom{abc}d}=(F_{abC}^{\phantom{abC}D}\bar{\epsilon}_{C'}^{\phantom{C'}D'}
+\bar{F}_{abC'}^{\phantom{abC'}D'}\epsilon_{C}^{\phantom{C}D})\sigma_{c}^{\phantom{c}CC'}
\sigma^d_{\phantom{d}DD'}.
\label{eq:decompose-riemann} 
\end{equation}

In the important particular case of a torsion-free connection which is compatible with the metric (Levi-Civita connection) the curvature spinors gain further symmetries. These are
\begin{eqnarray*}
&& X_{(AB)CD}=X_{ABCD}\;,\quad
X_{ABCD}=X_{CDAB}\;,\\
&& \Phi_{(AB)C'D'}=\Phi_{ABC'D'}. 
\end{eqnarray*}
Given these additional symmetries we find that the spinor $\Phi_{ABC'D'}$ is already in its 
irreducible form and it is called the Ricci spinor. The irreducible decomposition of the spinor $X_{ABCD}$ yields 
\begin{equation}
 X_{ABCD}=\Psi_{ABCD}+\Lambda(\epsilon_{AD}\epsilon_{BC}+\epsilon_{AC}\epsilon_{BD}),
\end{equation}
where $\Psi_{ABCD}$ is a totally symmetric spinor called the {\em Weyl spinor} and $\Lambda$ is related to the scalar curvature by the formula $\Lambda=R/24$. The curvature spinors and the torsion spinor are defined up to a constant scalar factor.

When working with a spin covariant derivative $D_a$ it is convenient to introduce the differential operator
\begin{equation}
D_{AA'}\equiv\sigma^a_{\phantom{a}AA'}D_a\;, 
\label{eq:d-2-index}
\end{equation}
which enables us to render any expression containing spin covariant derivatives as 
an expression containing only spinor indices. Also the commutation $D_{AA'}D_{BB'}-D_{BB'}D_{AA'}$ 
can be formally decomposed into irreducible parts as follows
\begin{equation}
D_{AA'}D_{BB'}-D_{BB'}D_{AA'}=\overline{\epsilon}_{A'B'}\DAl_{AB}+\epsilon_{AB}\DAl_{A'B'},
\label{eq:covd-box}
\end{equation}
where 
\begin{equation}
\DAl_{AB}\equiv D_{(A|A'}D^{A'}_{\phantom{A'}|B)}\;,\quad
\DAl_{A'B'}\equiv D_{A(A'}D^{A}_{\phantom{A}B')}\;, 
\end{equation}
are linear differential operators. 
The action of these operators on a spinor of any rank is obtained 
from the spinor expression of the Ricci identity of $D_a$ and 
the expression of the Riemann tensor in terms of the curvature spinors. 
The results for the case of a rank-1 spinor are 
\begin{equation}
\DAl_{BC}\xi^D=X_{D\phantom{A}BC}^{\phantom{D}A}\xi^D-
\Omega^{A'D}_{\phantom{A'D}BC}D_{DA'}\xi^A
\label{eq:expand-box}
\end{equation}

\section{The package {\em Spinors} and its relation to {\em xAct}}
\label{sec:xact}
The package {\em Spinors} is a {\em Mathematica} package which implements the spinor calculus as described in the previous section. {\em Spinors} is part of {\em xAct} \cite{XACT}, 
which is a system to do tensor analysis by computer written mostly in the {\em Mathematica} programming language with a smaller part in C. The composition is roughly 16000 lines of {\em Mathematica} code and 2700 lines of C code. As of October 2011 the version of {\em xAct} is 1.0.3. and the complete system is free software available under the terms of the GPL license.  

\begin{figure}[h]
 \begin{center}
\includegraphics[width=\columnwidth]{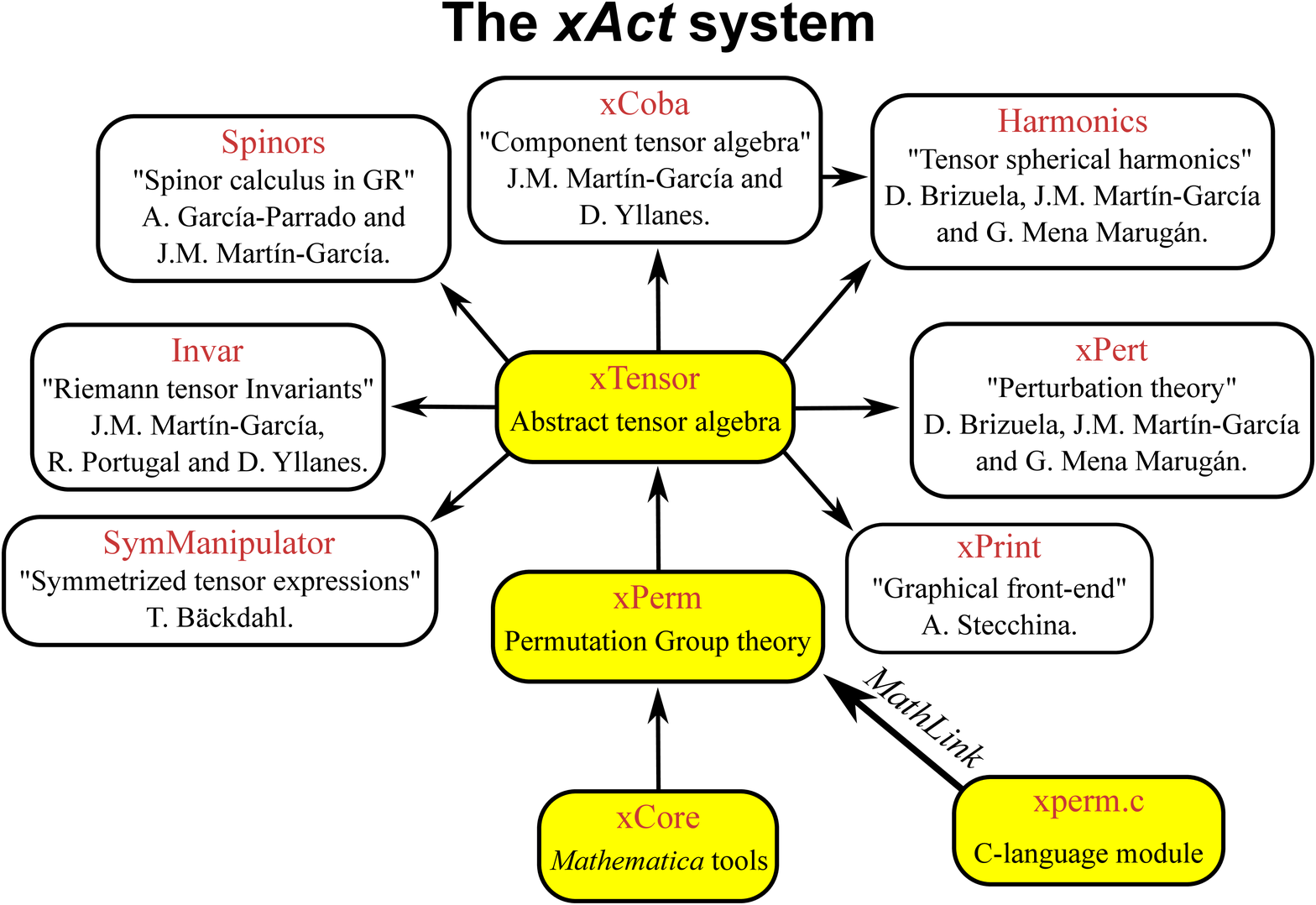}
\end{center}
\caption{\label{fig:xact} Graphical depiction of the xAct packages and their dependencies. 
The packages coloured in yellow can be regarded as the system Kernel and have been developed by 
J.M. Mart\'{\i}n-Garc\'{\i}a.}
\end{figure}

The system {\em xAct} is organised as a suite of inter-dependent {\em Mathematica} packages  
which can be regarded as different-purpose modules loadable on-demand. The packages and the relations 
among them are depicted in figure \ref{fig:xact}. The arrows indicate which packages of the suite 
a given package relies upon and we can see that there are three packages 
({\em xCore}, {\em xPerm} and {\em xTensor}) which act as kernel for the whole implementation. 
This means that these packages yield the basic framework to set up 
any computation requiring tensor analysis. In addition the module {\em xperm.c} is a piece of code in 
C language devised to speed-up the group theoretical computations needed to {\em canonicalize} 
tensor expressions \cite{XPERM}. When loading any other package of {\em xAct}, all the necessary 
packages are loaded sequentially, following the order shown in figure \ref{fig:xact}. 

The {\em xAct} system can be enlarged  by adding new packages for specific purposes, 
as long as the dependencies just described are kept.    
Some of the packages shown in figure \ref{fig:xact} have been already described in the literature. 
This is the case of {\em xPerm} \cite{XPERM}, {\em Invar} \cite{INVAR1,INVAR2}
and {\em xPert} \cite{XPERT}. In addition to these references, each package included in 
{\em xAct} has a documentation file, which explains its main features and contains a tutorial.

The {\em xAct} system provides support to define and work with general vector bundles
on any manifold and dimension. This means that the main features of spinor algebra
and spinor calculus are generically built into {\em xAct} for any dimension. However, in the
particular case of Lorentzian geometry in dimension four, extra features arise as described in section
\ref{sec:preliminaries} and hence it was necessary to develop a new package to take care 
of these features. For example the presence of the antisymmetric metric $\epsilon_{AB}$
forces us to be careful with the conventions to raise and lower indices. Also 
we need to decide which index configuration represents the {\em basic} Kronecker delta and 
which index configuration represents a derived quantity. In our case all these conventions were
laid in eqs. (\ref{eq:raise-lower})-(\ref{eq:kronecker-delta}) and one needs to take these 
particular conventions to the software implementation.  Another delicate task is the 2-index
representation of the spin covariant derivatives and their properties (eqs. (\ref{eq:d-2-index})-
(\ref{eq:covd-box})). This representation only occurs in the spinor calculus and hence we had to code 
the corresponding properties explicitly for the {\em Spinors} package. Finally the formulae relating spinors and 
tensors had to be studied and coded from scratch 
(this was perhaps one of the most time-consuming 
tasks in the development of {\em Spinors}). The conclusion of all of this is that spinor theory 
is complex enough to develop a new package for the {\em xAct} suite. 

The package spinors has already been used by a number of authors in their research. 
An example of this is the invariant construction of {\em Kerr initial data} in 
\cite{KERR-INVARIANT-TJ,KERR-INVARIANT-PRL} (see also \cite{KERR-INVARIANT-PRS} for a generalisation
of these results). The present authors have also used {\em Spinors} in the investigation of 
the invariant properties of type D vacuum solutions of the Einstein's field equations \cite{EDGAR-D}. 

\section{Working with {\em Spinors}}
\label{sec:spinors-description}
Assuming {\em xAct} has been installed one 
loads {\em Spinors} in a {\em Mathematica} session by typing 

\begin{flushleft}
\Mathsize
\stepcounter{mathex}
\MathIn{mathex}{<< xAct\`{}Spinors\`{}} 
{\tt 
------------------------------------------------------------

Package xAct`Spinors`  version 1.0.2, \{2011,10,25\}

CopyRight (C) 2006-2011, Alfonso Garcia-Parrado Gomez-Lobo and Jose M. Martin-Garcia, under the General Public License.
------------------------------------------------------------
}
\end{flushleft}

\noindent
This will load the package {\em Spinors} together with the other packages of the {\em xAct} suite which {\em Spinors} relies on. These are {\em xCore}, {\em xPerm} and 
{\em xTensor} 
(see section \ref{sec:xact} and figure \ref{fig:xact}). 
In this work we will only explain the features of these packages which are required for our implementation and
we refer the reader to their documentation for further details. 

Next we need to declare a 4-dimensional Lorentzian manifold by means of the standard  
{\em xAct} machinery:

\begin{flushleft}
\Mathsize
\stepcounter{mathex}
\MathIn{mathex}{DefManifold[M4, 4, \{a, b, c, d, f, h, p\}]}
\stepcounter{mathex}
\MathIn{mathex}{DefMetric[\{1,3,0\}, g[-a,-b], CD]}
\end{flushleft}

\noindent
The list \icode{\{a, b, c, d, f, h, p\}} corresponds to the space-time abstract
indices which will be used in tensor expressions and the list \icode{\{1,3,0\}}
in \icode{DefMetric} serves to indicate the canonical form of the metric tensor
\icode{g} (its canonical form contains once +1, three times -1 and zero times 0,
thus it corresponds to a Lorentzian metric). The symbol \icode{CD} represents
the Levi-Civita connection compatible with the metric \icode{g} and in addition
a number of quantities (the Riemann tensor, the Ricci tensor, the Weyl tensor,
etc) are also created automatically after issuing the command \icode{DefMetric}.

So far we have used commands belonging to {\em xTensor} and we have now the
set-up necessary to start working with {\em Spinors}. The first step is the
introduction of a {\em spin structure}. This is achieved as follows 

\begin{flushleft}
\Mathsize
\stepcounter{mathex}
\MathIn{mathex}{DefSpinStructure[g, Spin, \{A, B, C, D, F, P\}, $\epsilon$, $\sigma$, CDe]}
\end{flushleft}

\noindent
Several new objects are defined alongside this command. These are the spin
bundle \icode{Spin}, its abstract indices \icode{\{A, B, C, D, F, P\}},
the spin metric \icode{$\epsilon$}, the soldering form \icode{$\sigma$}
and the spin covariant derivative \icode{CDe} compatible with both the
space-time metric \icode{g} and the spin metric \icode{$\epsilon$}.
The spin bundle \icode{Spin\dd} together with 
its structures and the curvature spinors are automatically defined with 
this command. 
For example the Weyl and Ricci spinors are 

\begin{flushleft} 
\Mathsize
\stepcounter{mathex}
\MathIn{mathex}{\{PsiCDe[-A, -B, -C, -D], PhiCDe[-A, -B, -C\dd, -D\dd\;]\}}
\MathOut{mathex} $\{\Psi[\nabla]_{\rm A B C D},\;\Phi[\nabla]_{\rm A B C\dd D\dd}\}$
\end{flushleft}

Additional options controlling the displayed form of the different quantitites
automatically defined can be supplied to \icode{DefSpinStructure}.
For example we may add the options 

\vspace{.4cm} 
{ 
\Mathsize
{\tt SpinorPrefix-> SP,}\ \  {\tt SpinorMark-> "S"}.
\vspace{.4cm} 
}

\noindent
The symbol \icode{SP} will be prepended to the tensor (spinor) counterpart of
any spinor (tensor) and the string \icode{"S"} will be used in the displayed
representation (see below for explicit examples).
From now on it will be understood that these options were used in the 
command \icode{DefSpinStructure} above.

Primed indices are entered with the ``dagger'' symbol \icode{\ddn} 
(entered via the keyboard shortcut \frame{esc}-dg-\frame{esc}). 
Take now the Ricci tensor \icode{RicciCD[-a, -b]}. 
Its spinor counterpart is represented by 
\icode{SP\!\link RicciCD[-A, -A\dd, -B, -B\dd]}
where \icode{SP} was the tag declared through the option \icode{SpinorPrefix}
in \icode{DefSpinStructure}. Again, this tag is used to construct the symbol
defining the spinor counterpart of any tensorial quantity, in the way
illustrated by this example. In the mathematica notebook this is  

\begin{flushleft}
\Mathsize
\stepcounter{mathex}
\MathIn{mathex}{SP\!\link RicciCD[-A, -A\dd, -B, -B\dd]}
\MathOut{mathex} $S{\rm R}[\nabla]_{\rm A A\dd B B\dd}$
\end{flushleft}

\noindent
The linking symbol ``\link'' (entered through
the key combinations \frame{esc}-u[-\frame{esc}) serves to link the symbol
chosen to represent the spinor prefix with the symbol representing the tensor.
This linking symbol is stored in the variable \icode{\$LinkCharacter} which
can be freely modified.
In addition to this, it is possible to modify the default printing options and
obtain outputs similar to the formulae described in section
\ref{sec:preliminaries}. This is done as follows for the primed indices

\begin{flushleft}
\Mathsize
\stepcounter{mathex}
\MathIn{mathex}{
PrintAs[A\dd] ${}^{\wedge}$="A$'$";
PrintAs[B\dd] ${}^{\wedge}$="B$'$"; \\
PrintAs[C\dd] ${}^{\wedge}$="C$'$";
PrintAs[D\dd] ${}^{\wedge}$="D$'$";
PrintAs[F\dd] ${}^{\wedge}$="F$'$";
}
\end{flushleft}

\noindent
Also we can modify the printing output of any tensor or spinor

\begin{flushleft}
\Mathsize
\stepcounter{mathex}
\MathIn{mathex}{PrintAs[$\epsilon\dd$] ${}^{\wedge}$= "$\bar{\epsilon}$";}
\end{flushleft}

The command \icode{Decomposition} can be used to find the decomposition into
irreducible parts of any other curvature spinor (it is possible to indicate the
curvature spinor being decomposed as an additional argument). For example

\begin{flushleft} 
\Mathsize
\stepcounter{mathex}
\MathIn{mathex}{SP\!\link RiemannCD[-A, -A\dd, -B, -B\dd,-C, -C\dd, -D, -D\dd]\\ // Decomposition}
\MathOut{mathex}
$\Psi[\triangledown]\dd_{\rm A'B'D'C'} \epsilon_{\rm AB}\epsilon_{\rm CD} -
\Phi[\triangledown]_{\rm ABD'C'} \epsilon_{\rm DC}
\bar{\epsilon }_{\rm A'B'} - 2\Lambda[\triangledown](\epsilon_{\rm AD} 
\epsilon_{\rm BC} \bar{\epsilon }_{\rm A'D'}
\bar{\epsilon }_{\rm B'C'} -  \epsilon_{\rm AC} \epsilon_{\rm BD} \
\bar{\epsilon }_{\rm A'C'} \bar{\epsilon }_{\rm B'D'}) -  \
\Phi[\triangledown]_{\rm DCA'B'} \epsilon_{\rm AB} \
\bar{\epsilon }_{\rm D'C'} + \Psi[\triangledown]_{\rm ABDC}
\bar{\epsilon }_{\rm A'B'} \bar{\epsilon }_{\rm C'D'}$
\end{flushleft}

\noindent
The spin covariant derivative \icode{CDe} can be handled as a 2-index covariant
derivative ...

\begin{flushleft}
\Mathsize
\stepcounter{mathex}
\MathIn{mathex}{CDe[-F,-F\dd]@PsiCDe[-A,-B,-C,-D]}
\MathOut{mathex} $\nabla_{\rm F F'}\Psi[\nabla]_{\rm ABCD}$
\end{flushleft}

\noindent
.. or as a covariant derivative in a vector bundle 

\begin{flushleft}
\Mathsize
\stepcounter{mathex}
\MathIn{mathex}{SeparateSolderingForm[][\%]}
\MathOut{mathex} $\sigma^{\rm a}{}_{\rm FF'} 
\triangledown_{\rm a}\Psi[\triangledown]_{\rm ABCD}$.
\end{flushleft}
\noindent
The command \icode{SeparateSolderingForm} enables us to transform spinor
indices into tensor ones (cf. \S\ref{subsec:tensor-spinor} for more details
about this). 

Spinors are defined by means of \icode{DefSpinor}, which is just a special
call to the {\em xAct} command \icode{DefTensor} (by default the option
\icode{Dagger->Complex} is assumed).

\begin{flushleft}
\Mathsize
\stepcounter{mathex}
\MathIn{mathex}{DefSpinor[$\kappa$[-A, -A\dd], M4]}
\DefTensorMessage{$\kappa$[-A, -A\dd]}
\DefTensorMessage{$\kappa$\dd[-A\dd, -A]}
\end{flushleft}
If one wishes to work with Hermitian spinors then this is done by using the
option \icode{Dagger -> Hermitian} on the command \icode{DefSpinor}. Under this
assumption one has that \icode{$\kappa$[-A,-A\dd]} is invariant under complex
conjugation

\begin{flushleft}
\Mathsize
\stepcounter{mathex}
\MathIn{mathex}{$\kappa$[-A,-A\dd] // Dagger // InputForm}
\MathOut{mathex} {\tt $\kappa$[-A,-A\dd]} 
\end{flushleft}

The canonicalizer of {\em xAct} is \icode{ToCanonical} and it can deal with the
canonicalization of spinor expressions without any additional user input. 
It is beyond the present article to explain the workings of the canonicalization
procedure and the reader is referred to \cite{XPERM} and the {\em xTensor}
documentation for additional details. Similarly, the {\em xAct} command
\icode{ContractMetric} takes care automatically of the conventions for
raising and lowering indices in spinor expressions.  
We present next some explicit examples about these issues 

\begin{flushleft}
\Mathsize
\stepcounter{mathex}
\MathIn{mathex}{DefSpinor[$\mu$[A], M4]}
\DefTensorMessage{$\mu$[A]}
\DefTensorMessage{$\mu$\dd[A\dd]} 
\end{flushleft}

\begin{flushleft}
\Mathsize
\stepcounter{mathex}
\MathIn{mathex}{
\{$\mu$[-B]$\epsilon$[A,B],$\mu$[B]$\epsilon$[-B,-A],$\epsilon$\dd[C\dd,A\dd]
CDe[-A,-A\dd]@$\mu$[-B]\}}
\MathOut{mathex} {$\{\epsilon^{\rm AB} \mu_{\rm B},\ \epsilon_{\rm BA} \mu^{\rm B},\ 
\bar{\epsilon}^{\rm C'A'} \triangledown_{\rm AA'}\mu_{\rm B}\}$} 
\end{flushleft}

\begin{flushleft}
\Mathsize
\stepcounter{mathex}
\MathIn{mathex}{ContractMetric[\%]}
\MathOut{mathex} {$\{\mu^{\rm A},\ \mu_{\rm A},\ \triangledown_{\rm A}{}^{\rm C'}\mu_{\rm B}\}$}
\end{flushleft}

\begin{flushleft}
\Mathsize
\stepcounter{mathex}
\MathIn{mathex}{\{
$\mu$[A]$\mu$[-A],
$\kappa$[-A, -B\dd]$\mu$[B\dd],
CDe[-F, -F\dd]@CDe[-A, F\dd]@PsiCDe[-B, -C, -D, -P]+
CDe[-F, F\dd]@CDe[-A, -F\dd]@PsiCDe[-B, -C, -D,-P]
\}}
\MathOut{mathex} { 
$\{\mu_{\rm A} \mu^{\rm A},\ \kappa_{\rm AB'}\bar\mu^{\rm B'},\ \triangledown_{\rm FF'}
\triangledown_{\rm A}{}^{\rm F'}\Psi[\triangledown]_{\rm BCDP} + \triangledown_{\rm F}{}^{\rm F'}\triangledown_{\rm AF'}
\Psi[\triangledown]_{\rm BCDP}\}$} 
\end{flushleft}

\begin{flushleft}
\Mathsize
\stepcounter{mathex}
\MathIn{mathex}{ToCanonical[\%]}
\MathOut{mathex} {$\{0,\ - \kappa_{\rm A}{}^{\rm B'}\bar\mu_{\rm B'},\ 0\}$} 
\end{flushleft}

Also the commutation of the spin covariant derivatives shown in (\ref{eq:covd-box})
is implemented

\begin{flushleft}
\Mathsize
\stepcounter{mathex}
\MathIn{mathex}{CDe[-C,-C\dd]@CDe[-B,-B\dd]@$
\kappa$[-A,-A\dd]-\\
CDe[-B,-B\dd]@CDe[-C,-C\dd]@$\kappa$[-A,-A\dd]}
\MathOut{mathex} {- $\triangledown_{\rm BB'}\triangledown_{\rm CC'}
\kappa_{\rm AA'} + \triangledown_{\rm CC'}\triangledown_{\rm BB'}\kappa_{\rm AA'}$}
\end{flushleft}

\begin{flushleft}
\Mathsize
\stepcounter{mathex}
\MathIn{mathex}{SortSpinCovDs[\%,CDe]}
\MathOut{mathex} {
$\bar\epsilon_{\rm C'B'}(\square[\triangledown]_{\rm CB}\kappa_{\rm AA'}) + 
\epsilon_{\rm CB}(\square[\triangledown]_{\rm C'B'}\kappa_{\rm AA'})$}
\end{flushleft}

The differential operator $\DAl_{AB}$ can be written in terms of the spin 2-index covariant derivatives as illustrated in the following example

\begin{flushleft}
\Mathsize
\stepcounter{mathex}
\MathIn{mathex}{BoxCDe[-A,-B]@$\kappa$[-C,-C\dd]}
\MathOut{mathex} $\square[\nabla]_{\rm AB}\kappa_{\rm CC'}$\\
\end{flushleft}

\begin{flushleft}
\Mathsize
\stepcounter{mathex}
\MathIn{mathex}{BoxToCovD[\%,BoxCDe]}
\MathOut{mathex}
$\frac{1}{2} (\nabla_{\rm AA'}\nabla_{\rm B}{}^{\rm A'}\kappa_{\rm CC'} + \nabla_{\rm BA'}
\nabla_{\rm A}{}^{\rm A'}\kappa_{\rm CC'})$
\end{flushleft}

\begin{flushleft}
\Mathsize
\stepcounter{mathex}
\MathIn{mathex}{BoxToCurvature[\%\%,BoxCDe]}
\MathOut{mathex} {- $\Phi[\nabla]_{\rm ABA' C'}\kappa_{\rm C}{}^{\rm A'} -  X[\nabla]_{\rm DCAB}\kappa^{\rm D}{}_{\rm C'}$}
\end{flushleft}

In our previous examples we worked with the spin covariant derivative arising from the 
Levi-Civita connection but we can introduce other arbitrary spin covariant derivatives. For example
 
\begin{flushleft}
\Mathsize
\stepcounter{mathex}
\MathIn{mathex}{DefSpinCovD[nb[-a], $\sigma$, 
SymbolOfCovD-> \{"|","D"\}, Torsion->True]}
\end{flushleft}
As we see in the example the command \icode{DefSpinCovD} shares some
similarities with the {\em xTensor} command \icode{DefCovD}. In addition to
the spin covariant derivative output symbols, we also need to specify the spin
structure which the spin covariant derivative is compatible with
(this is $\sigma$ in our example). Hence

\begin{flushleft}
\Mathsize
\stepcounter{mathex}
\MathIn{mathex}{\{nb[-a]@$\sigma$[b, -A, -A\dd],\quad 
nb[-a]@$\sigma$[-b, A, A\dd] \}}
\MathOut{mathex} {\{0,\quad 0\}}
\end{flushleft}

Also, a number of quantities are automatically defined in addition to the spin
covariant derivative \icode{nb}. Since we used the option \icode{Torsion->True}
the torsion is among those and one can work with both its tensor and spinor
forms.
 
\begin{flushleft}
\Mathsize
\stepcounter{mathex}
\MathIn{mathex}{SeparateSolderingForm[]@Torsionnb[a,-b,-c]}
\MathOut{mathex} {${\rm ST}[{\rm D}]^{\rm AA'}{}_{\rm BB'CC'} 
\sigma^{\rm a}{}_{\rm AA'}\sigma_{\rm b}{}^{\rm BB'} \sigma_{\rm c}{}^{\rm CC'}$}
\end{flushleft}

\begin{flushleft}
\Mathsize
\stepcounter{mathex}
\MathIn{mathex}{PutSolderingForm@Decomposition@\%}
\MathOut{mathex} {$\epsilon^{\rm AD}\bar\epsilon^{\rm A'D'} 
(\Omega[{\rm D}]\dd_{\rm DD'B'C'} \epsilon_{\rm BC} + \Omega[{\rm D}]_{\rm D'DBC} \bar\epsilon_{\rm B'C'})$}
\end{flushleft}

\begin{flushleft}
\Mathsize
\stepcounter{mathex}
\MathIn{mathex}{ContractMetric@\%}
\MathOut{mathex} {
$\Omega[{\rm D}]\dd^{\rm AA'}{}_{\rm B'C'}\epsilon_{\rm BC} + 
\Omega[{\rm D}]^{\rm A'A}{}_{\rm BC}\bar\epsilon_{\rm B'C'}$}
\end{flushleft}

The first step finds the relation between the torsion tensor and the torsion
spinor and the second step computes its irreduccible decomposition according
to eq.  (\ref{eq:torsion-decomposition}). Any spin covariant derivative can be
represented in single index and two-index notation.

\begin{flushleft}
\Mathsize
\stepcounter{mathex}
\MathIn{mathex}{nb[-A,-A\dd]@$\mu$[-C]}
\MathOut{mathex} {$D_{\rm AA'}\mu_{\rm C}$}
\end{flushleft}

\begin{flushleft}
\Mathsize
\stepcounter{mathex}
\MathIn{mathex}{SeparateSolderingForm[][\%,nb]}
\MathOut{mathex} {$\sigma^{\rm a}{}_{\rm AA'} D_{\rm a}\mu_{\rm C}$}
\end{flushleft}

Finally we remark that it is possible to define a spin structure for a metric
connection with torsion. In this case \icode{DefSpinStructure} defines the
torsion spinors automatically.
 
\subsection{Relations between tensors and spinors}
\label{subsec:tensor-spinor}

One of the strongest points of {\em Spinors} is its ability 
to transform tensor expressions into spinor ones and back. 
The transformation rules are ilustrated by (\ref{ivv})-(\ref{vvi}) and 
to work out these expressions in explicit examples we need to repeatedly use (\ref{eq:tetra-g}). 
To illustrate how this works in {\em Spinors} let us consider the following example: 
suppose that we have the Riemann tensor associated to the Levi-Civita connection 
and we wish to find its spinor form by following (\ref{ivv}). The procedure is then
  
\begin{flushleft}
\Mathsize
\stepcounter{mathex}
\MathIn{mathex}{PutSolderingForm@RiemannCD[-a,-b,-c,-d]}
\MathOut{mathex} {
${\rm R}[\triangledown]_{\rm abcd}\sigma^{\rm a}{}_{\rm AA'}\sigma^{\rm b}{}_{\rm BB'}
\sigma^{\rm c}{}_{\rm CC'}\sigma^{\rm d}{}_{\rm DD'}$}
\end{flushleft}

\begin{flushleft}
\Mathsize
\stepcounter{mathex}
\MathIn{mathex}{ContractSolderingForm@\%}
\MathOut{mathex} {${\rm SR}[\triangledown]_{\rm AA'BB'CC'DD'}$}
\end{flushleft}

The Riemann spinor can be transformed back into a tensor as follows

\begin{flushleft}
\Mathsize
\stepcounter{mathex}
\MathIn{mathex}{SeparateSolderingForm[$\sigma$][\%]}
\MathOut{mathex} {${\rm R}[\triangledown]_{\rm abcd}\sigma^{\rm a}{}_{\rm AA'}\sigma^{\rm b}{}_{\rm BB'}\sigma^{\rm c}{}_{\rm CC'}\sigma^{\rm d}{}_{\rm DD'}$}
\end{flushleft}

\begin{flushleft}
\Mathsize
\stepcounter{mathex}
\MathIn{mathex}{PutSolderingForm@\%}
\MathOut{mathex} {${\rm R}[\triangledown]_{\rm abcd}$}
\end{flushleft}
 
As we see in this simple example a tensor (resp. a spinor) is transformed into a spinor (resp. a tensor) by contracting it with a number of soldering forms in the appropriate way. The insertion of soldering forms is achieved with the command \icode{PutSolderingForm} and the elimination of their dummy indices with \icode{ContractSolderingForm}. 
The command \icode{SeparateSolderingForm} also inserts a number of soldering forms but unlike \icode{PutSolderingForm}, the expression on which it acts (a tensor or a spinor) is automatically replaced by its tensor or spinor counterpart. If the tensor or spinor counterpart has not been previously defined, then it is created automatically. Example:

\begin{flushleft}
\Mathsize
\stepcounter{mathex}
\MathIn{mathex}{DefTensor[M[-a, -b], M4]}
\DefTensorMessage{M[-a, -b]}
\end{flushleft}

\begin{flushleft}
\Mathsize
\stepcounter{mathex}
\MathIn{mathex}{SeparateSolderingForm[]@M[-a, -b]}
{\tt \textcolor{red}{SpinorOfTensor::name: Spinor of M not defined. Prepending SP.}}\\
\DefTensorMessage{SP\!\link M[-Q\$3,-Q\dd\$3,-Q\$5,-Q\dd\$5]}
\DefTensorMessage{SP\!\link M\dd[-Q\dd\$3,-Q\$3,-Q\dd\$5,-Q\$5]}
\MathOut{mathex}
{\tt ${\rm SM}_{\rm AA'BB'} \sigma_{\rm a}{}^{\rm AA'}
\sigma_{\rm b}{}^{\rm BB'}$}
\end{flushleft}

All these commands admit a number of options to select the tensor (spinor) indices on which one wants to act and what tensor (spinor) indices are going to be contracted. We refer the reader to the on-line documentation of each command for the complete list of available options. Another related  possibility also covered is the case in which one has a tensor (resp. spinor) already defined in the session and one wishes to introduce its spinor (resp. tensor) counterpart. The way in which this is achieved is through the command \icode{DefSpinorOfTensor} (resp. 
\icode{DefTensorOfSpinor}). These commands allow the user to choose the symbols representing the tensor or spinor counterparts. For example take the tensor 
\icode{M[-a,-b]} defined above and suppose that we have not used the automatic procedure to 
define its spinor counterpart described above. Then one can do the following 

\begin{flushleft}
\Mathsize
\stepcounter{mathex}
\MathIn{mathex}{ 
DefSpinorOfTensor[SPM[-A,-A\dd,-B,-B\dd], M[-a,-b], $\sigma$, PrintAs 
$\rightarrow\mbox{"}\mathcal{M}\mbox{"}$]}
\DefTensorMessage{SPM[-A,-A\dd,-B,-B\dd]}
\DefTensorMessage{SPM\dd[-A\dd,-A,-B\dd,-B]}
\end{flushleft}

The tensor \icode{M} and the spinor \icode{SPM} are now paired to each 
other. For example, if we act on \icode{M} with \icode{SeparateSolderingForm[]} the system will use the spinor which we have defined above rather than an automatic definition.

\begin{flushleft}
\Mathsize
\stepcounter{mathex}
\MathIn{mathex}{SeparateSolderingForm[]@M[-a,-b]}
\MathOut{mathex}
{\tt ${\mathcal M}_{\rm AA'BB'} \sigma_{\rm a}{}^{\rm AA'} 
\sigma_{\rm b}{}^{\rm BB'}$}
\end{flushleft}
If the tensor \icode{M} had had any symmetry, then the symmetries of the spinor \icode{SPM}  would have been automatically computed. 

When translating spinor into tensor expressions it is important to control how products of soldering forms transform into tensors. A simple example of such a transformation is shown in eq. (\ref{eq:sigma-product}) and products of soldering forms with more factors will arise when transforming complicated spinor expressions into tensor ones. The way of computing products of soldering forms in {\em Spinors} is through the command \icode{ContractSolderingForm}.
For example, the simplest case is the product of two soldering forms
\begin{flushleft}
\Mathsize
\stepcounter{mathex}
\MathIn{mathex}{$\sigma$[a,-A,-A\dd] $\sigma$[b,-B,A\dd]//ContractSolderingForm}
\MathOut{mathex} {$\Sigma\sigma^{\rm ab}{}_{\rm AB}$}
\end{flushleft}
The mixed quantity $\Sigma\sigma^{\rm ab}{}_{\rm AB}$ is entered through the keyboard 
as \icode{Sigma$\sigma$[a, b, -A, -B]} and its square results in the tensor $G_{abcd}$ introduced in eq. (\ref{eq:tetra-g}). The tensor $G_{abcd}$ will be referred to as the {\em tetra-metric} and it is one of the quantities automatically defined by 
\icode{DefMetric} when the manifold dimension is four. In this way, if the metric name symbol is \icode{g} then $G_{abcd}$ is represented by the symbol 
\icode{Tetrag} and eq. (\ref{eq:tetra-g}) by the rule \icode{TetraRule[g]}

\begin{flushleft}
\Mathsize
\stepcounter{mathex}
\MathIn{mathex}{Tetrag[-a, -b, -c, -d]}
\MathOut{mathex} {$\rm{Gg_{abcd}}$}\\
\end{flushleft}

\begin{flushleft}
\Mathsize
\stepcounter{mathex}
\MathIn{mathex}{\% /. TetraRule[g]}
\MathOut{mathex} {$\rm{\frac{1}{2}i 
\epsilon g_{abcd} + \frac{1}{2} g_{ad} g_{bc} - \
\frac{1}{2} g_{ac} g_{bd} + \frac{1}{2} g_{ab} g_{cd}}$}
\end{flushleft}

\noindent
The square of $\Sigma\sigma^{\rm ab}{}_{\rm AB}$ is always automatically
replaced by the tetra-metric

\begin{flushleft}
\Mathsize
\stepcounter{mathex}
\MathIn{mathex}{Sigma$\sigma$[-a, -b, -A, -B]
Sigma$\sigma$[-c, -d, A, B]}
\MathOut{mathex} {$\rm{Gg_{dbac}}$ }
\end{flushleft}

\noindent
The main interest of the tetra-metric is that any contracted product of
soldering forms with no free spinor indices can be always expressed as a
product of tetra-metrics. This is precisely the kind of product which arises
naturally when translating spinor expressions into tensor ones and back.
Consider the following example: if $\Psi_{ABCD}$ is the Weyl spinor, we wish
to write the scalar quantity $\Psi_{ABCD}\Psi^{ABCD}$ as an expression in terms
of the Weyl tensor. The Weyl spinor and the Weyl tensor $W_{abcd}$ are related
through the relation 
\begin{equation}
\Psi_{ABCD} = \frac{1}{4} W_{abcd} \
\sigma^{a}{}_{A}{}^{A'} \sigma^{b}{}_{BA'} \sigma^{c}{}_{C}{}^{C'} \
\sigma^{d}{}_{DC'}\;,
\label{eq:weyl-spinor-to-tensor}
\end{equation}
and hence the scalar $\Psi_{ABCD}\Psi^{ABCD}$ can be computed by replacing 
the Weyl spinor according to (\ref{eq:weyl-spinor-to-tensor}). Equation 
(\ref{eq:weyl-spinor-to-tensor}) can be written as a {\em xAct} rule in the
following way
\begin{flushleft}
\Mathsize
\stepcounter{mathex}
\MathIn{mathex}{WSToWT=
IndexRule[PsiCDe[A\_, B\_, C\_, D\_],\\
1/4WeylCD[-a, -b, -c, -d]$\sigma$[a, A, A\dd]$\sigma$[b, B, -A\dd]\\
$\sigma$[c, C, C\dd]$\sigma$[d, D, -C\dd]]}
\MathOut{mathex} {\tt HoldPattern[$\Psi[\nabla]^{\underline{ABCD}}$]:
\!\!\!\!\!\!{\tiny $\rightarrow$}\newline 
Module[\{a, A\dd, b, c, C\dd, d\}, 
$\frac{1}{4} W[\triangledown ]_{\rm abcd} \sigma^{\rm aA}{}^{\rm A'} \
\sigma^{\rm bB}{}_{\rm A'} \sigma^{\rm cC}{}^{\rm C'} \sigma^{\rm dD}{}_{\rm C'}$] }
\end{flushleft}
This construct is called in {\em xAct} an {\em index rule}. Its diffe\-ren\-ce with a {\em Mathematica} (delayed) rule is that dummy indices can be included in the right hand side of the index rule without caring about the collision of these indices with other dummy indices already present in the expression in which the replacement is being done. Dummy indices will be automatically re-named to avoid any index collision. The reader is referred to the {\em xAct} documentation for further details about this. 

We can use now the rule defined above to find the tensor expression of any scalar invariant written in terms of the Weyl spinor. In our example the actual computation runs as follows
\begin{flushleft}
\Mathsize
\stepcounter{mathex}
\MathIn{mathex}{\\ PsiCDe[-A, -B, -C, -D] PsiCDe[A, B, C, D] /. WSToWT}
\MathOut{mathex} {  
$\frac{1}{16} W[\triangledown ]_{\rm acfj} W[\triangledown ]_{\rm bdhl} \
\sigma^{\rm a}{}_{\rm A}{}^{\rm A'} \sigma^{\rm bAB'} \sigma^{\rm c}{}_{\rm BA'} \
\sigma^{\rm dB}{}_{\rm B'} \sigma^{\rm f}{}_{\rm C}{}^{\rm C'} \sigma^{\rm hCD'} \
\sigma^{\rm j}{}_{\rm DC'} \sigma^{\rm lD}{}_{\rm D'}$}
\end{flushleft}
\begin{flushleft}
\Mathsize
\stepcounter{mathex}
\MathIn{mathex}{ContractSolderingForm[\%, IndicesOf[Spin]]}
\MathOut{mathex} {  
$\frac{1}{16} g^{\rm af} g^{\rm bm} Gg^{\rm dc}{}_{\rm a}{}^{\rm h} 
Gg^{\rm lj}{}_{\rm b}{}^{\rm n} 
W[\triangledown ]_{\rm cfjm} W[\triangledown ]_{\rm dhln}$}
\end{flushleft}

The option \icode{IndicesOf[Spin]} used in \icode{ContractSolderingForm}
indicates that only spinor (dummy) indices in the product of soldering form
have to be taken into account in the contraction. In this way the final result
does not contain any spinor index and it is thus a tensor expression as desired.
One can now use the \icode{TetraRule[g]} discussed before to transform the
tetra-metrics into ordinary metrics and epsilon symbols (volume elements). 

\begin{flushleft}
\Mathsize
\stepcounter{mathex}
\MathIn{mathex}{\% /. TetraRule[g]; }
\stepcounter{mathex}
\MathIn{mathex}{ToCanonical@ContractMetric@\% }
\MathOut{mathex} {
$\frac{1}{8} W[\triangledown ]_{\rm abcd} W[\triangledown ]^{\rm abcd} +
\frac{1}{16}{\rm i}\epsilon g_{\rm cdfh} W[\triangledown ]_{\rm ab}{}^{\rm fh} 
W[\triangledown ]^{\rm abcd}$}
\end{flushleft}

\section{Example: The Sparling identity}
\label{sec:sparling}
As a final exercise with {\em Spinors} we show how to use the software to
derive the {\em Sparling identity}. This identity has the same information as
the Einstein field equations and can be formulated either in tensor or spinor
form. The spinor form of the equation has its origins in Witten's proof of the
positive mass theorem \cite{WITTEN-MASS} (see also \cite{NESTER-FORM}) while
the tensor form was found by Sparling (see \cite{FRAUENDIENER-PSEUDO,SZABADOS-PSEUDO} 
for a geometric derivation of this form of the equation). 

The set-up is as follows: let $\lambda^A$ be any rank-1 spinor and define the
following quantity
\begin{equation}
\Xi_{AA'BB'}\equiv\frac{\rm i}{2}(\bar\lambda_{A'}\nabla_{BB'}\lambda_A-\bar\lambda_{B'}\nabla_{AA'}\lambda_B). 
\label{eq:nester-witten}
\end{equation}
The spinor $\Xi_{AA'BB'}$ is called the {\em Nester-Witten} spinor and it has
the algebraic property $\Xi_{AA'BB'}+\Xi_{BB'AA'}=0$. Hence, its tensor
counterpart, defined by
\begin{equation}
\mathcal{F}_{ab}\equiv \sigma_{a}^{\phantom{a}AA'}\sigma_{b}^{\phantom{b}BB'}\Xi_{AA'BB'} 
\end{equation}
is an antisymmetric tensor and it can be regarded as a 2-form.
 
\begin{theorem}
The 2-form $\mathcal{F}_{ab}$ fulfills the relation (Sparling identity)
\begin{equation}
3\nabla_{[a}{\mathcal F}_{bc]}= 
\eta_{abcf}\left(\mathcal{Z}_d^{\phantom{d}[df]}- \frac{1}{4} G_{d}{}^{f} \xi^{d}\right)\;, 
\label{eq:sparling-equation}
\end{equation}
where $\xi^a$ is the tensor representing the spinor $\lambda^A\bar\lambda^{A'}$,
$G_{ab}$ is the Einstein tensor, $\eta_{abcd}$ the volume 4-form (both with
respect to the space-time metric) and ${\mathcal Z}_{abc}$ is a tensor
fulfilling the ``dominant property'', namely for any three causal
future-directed null vectors $k_1^a$, $k_2^a$, $k_3^a$ one has the property
\begin{equation}
{\mathcal Z}_{abc}k_1^ak_2^bk_3^c\geq 0. 
\label{eq:dominant-property}
\end{equation}
\end{theorem}

\noindent
{\em Proof:} We carry out the proof of this result using the tools introduced
in section \ref{sec:spinors-description} (we work in the same {\em Spinors}
session as the one used in that section). First of all, we need to define the
spinors and tensors intervening in our problem

\begin{flushleft}
\Mathsize
\stepcounter{mathex}
\MathIn{mathex}{DefSpinor[$\lambda$[-A], M4]}
\DefTensorMessage{$\lambda$[-A]}
\DefTensorMessage{$\lambda$\dd[-A\dd]}
\end{flushleft}

\begin{flushleft}
\Mathsize
\stepcounter{mathex}
\MathIn{mathex}{PrintAs@$\lambda$\dd\ ${}^{\wedge}$= "$\overline\lambda$";}
\end{flushleft}

\begin{flushleft}
\Mathsize
\stepcounter{mathex}
\MathIn{mathex}{
DefSpinor[$\Xi$[-A, -A\dd, -B, -B\dd], M4, 
 GenSet[-Cycles[\{-A, -B\}, \{-A\dd, -B\dd\}]]]}
\DefTensorMessage{$\Xi$[-A,-A\dd,-B,-B\dd]}
\DefTensorMessage{$\Xi$\dd[-A\dd,-A,-B\dd,-B]}
\end{flushleft}

\begin{flushleft}
\Mathsize
\stepcounter{mathex}
\MathIn{mathex}{PrintAs@$\Xi$\dd\ ${}^{\wedge}$= "$\overline\Xi$";}
\end{flushleft}

\begin{flushleft}
\Mathsize
\stepcounter{mathex}
\MathIn{mathex}{\\ DefTensorOfSpinor[$\mathcal F$[-a, -b], 
$\Xi$[-A, -A\dd, -B, -B\dd], $\sigma$]}
\DefTensorMessage{$\mathcal F$[-a,-b]}
\DefTensorMessage{$\mathcal F$\dd[-a,-b]}
\end{flushleft}

\begin{flushleft}
\Mathsize
\stepcounter{mathex}
\MathIn{mathex}{PrintAs@$\mathcal{F}$\ ${}^{\wedge}$= "$\mathcal{F}$"; 
PrintAs@$\mathcal{F}$\dd\ ${}^{\wedge}$= "$\overline{\mathcal{F}}$";}
\end{flushleft}

\begin{flushleft}
\Mathsize
\stepcounter{mathex}
\MathIn{mathex}{DefTensor[$\xi$[-a], M4]}
\DefTensorMessage{$\xi$[-a]}
\end{flushleft}

We change the default formatting of the volume form \icode{epsilong}

\begin{flushleft}
\Mathsize
\stepcounter{mathex}
\MathIn{mathex}{PrintAs[epsilong]\ ${}^{\wedge}$= "$\eta$";}
\end{flushleft}

We introduce a short form for the {\em xTensor} command \icode{IndexSolve}
(see the documentation of {\em xTensor} for further details about
\icode{IndexSolve})

\begin{flushleft}
\Mathsize
\stepcounter{mathex}
\MathIn{mathex}{IndSV[expr\_Equal]:= IndexSolve[expr, First@expr];} 
\end{flushleft}

We also define the shortcut canonicalization function \icode{TC} (combination of
the {\em xAct} commands \icode{ContractMetric} and \icode{ToCanonical})  

\begin{flushleft}
\Mathsize
\stepcounter{mathex}
\MathIn{mathex}{TC[expr\_] := ToCanonical[ContractMetric[expr]];} 
\end{flushleft}

Also we define a function named \icode{EqualTimes} to multiply by a quantity
both sides of an equation and canonicalize the result in just one step.

\begin{flushleft}
\Mathsize
\stepcounter{mathex}
\MathIn{mathex}{EqualTimes[Equal[lhs\_, rhs\_], x\_] := \\
Equal[TC[x lhs], TC[x rhs]];} 
\end{flushleft}

With all these preparations we introduce the Nester-Witten spinor definition,
as given by (\ref{eq:nester-witten}), in our {\em Spinors} session.

\begin{flushleft}
\Mathsize
\stepcounter{mathex}
\MathIn{mathex}{NesterWittenSpinor =\\ 
$\Xi$[-A, -A\dd, -B, -B\dd] == 1/2(-I $\lambda$\dd[-B\dd] 
CDe[-A, -A\dd]@$\lambda$[-B] + \\
I $\lambda$\dd[-A\dd] 
CDe[-B, -B\dd]@$\lambda$[-A])}
\MathOut{mathex}
{\tt $\Xi_{\rm AA'BB'} = \frac{1}{2} (-{\rm i} \overline{\lambda }_{\rm B'} 
\triangledown_{\rm AA'}\lambda_{\rm B} + {\rm i} \overline{\lambda }_{\rm A'} 
\triangledown_{\rm BB'}\lambda_{\rm A})$ }
\end{flushleft}

Our aim is to compute the quantity

\begin{flushleft}
\Mathsize
\stepcounter{mathex}
\MathIn{mathex}{d${\mathcal F}$ = 
CDe[-c]@${\mathcal F}$[-a, -b] // Antisymmetrize // TC}
\MathOut{mathex}
{\tt  $\frac{1}{3} \triangledown_{\rm a}\mathcal{F}_{\rm bc} - \frac{1}{3} 
\triangledown_{\rm b}\mathcal{F}_{\rm ac} + \frac{1}{3} 
\triangledown_{\rm c}\mathcal{F}_{\rm ab}$}
\end{flushleft}

We transform $\mathcal{F}_{ab}$ into the Nester-Witten spinor 
$\Xi_{AA'BB'}$ and then insert its explicit definition by means of the
{\em xAct} command \icode{IndexSolve}.

\begin{flushleft}
\Mathsize
\stepcounter{mathex}
\MathIn{mathex}{(d${\mathcal F}$ // SeparateSolderingForm[$\sigma$]) /. 
IndSV[NesterWittenSpinor];}
\end{flushleft}

The resulting expression (not shown due to lack of space) consists of 
second covariant spin derivatives of $\lambda_A$ and terms formed out of the
product $\nabla_{AA'}\lambda_B$, $\nabla_{AA'}\bar\lambda_{B'}$. The second 
covariant derivatives can be eliminated by means of the spinor Ricci identity
(\ref{eq:covd-box}) and (\ref{eq:expand-box}). In {\em Spinors} the procedure 
for doing this is as follows

\begin{flushleft}
\Mathsize
\stepcounter{mathex}
\MathIn{mathex}{SortSpinCovDs[\%, CDe];  }
\end{flushleft}

\begin{flushleft}
\Mathsize
\stepcounter{mathex}
\MathIn{mathex}{BoxToCurvature[\%, BoxCDe];}
\end{flushleft}

The arising curvature spinors have to be decomposed into irreducible parts 

\begin{flushleft}
\Mathsize
\stepcounter{mathex}
\MathIn{mathex}{d$\Xi$ = Decomposition[\%, Chi, CDe] // TC;}
\end{flushleft}

We split the previous equation into two parts:
terms containing covariant derivatives of $\lambda_{A}$
and terms which do not contain any covariant derivative. 

\begin{flushleft}
\Mathsize
\stepcounter{mathex}
\MathIn{mathex}{\{d$\Xi$1 = d$\Xi$ /.
 CDe[\_\!\_\!\_\!\_\;]@$\lambda$[\_\!\_\;]-> 0, 
 d$\Xi$2 = d$\Xi$ - d$\Xi$1\};}
\end{flushleft}

The aim is now to find the explicit tensor form of each part. 
The first part {\tt d$\Xi$1} is an expression which is linear in the curvature
spinors. We write the curvature spinors $\Phi_{ABA'B'}$ and $\Psi_{ABCD}$ in
terms of the trace-free Ricci tensor and the Weyl tensor respectively 
(the rule \icode{WSToWT} was defined from eq. (\ref{eq:weyl-spinor-to-tensor}),
see explanations coming after that equation.)

\begin{flushleft}
\Mathsize
\stepcounter{mathex}
\MathIn{mathex}{d$\Xi$1 /. WSToWT /. \\
IndexRule[PhiCDe[-A\_, -B\_, -A\dd\_, -B\dd\_], -1/2
TFRicciCD[-a, -b] $\sigma$[a, -A, -A\dd] 
$\sigma$[b, -B, -B\dd]] ;}
\end{flushleft}

Also we need to replace the products $\lambda_{A}\bar\lambda_{A'}$
by $\xi_a$.
\begin{flushleft}
\Mathsize
\stepcounter{mathex}
\MathIn{mathex}{\\ \% /. IndexRule[$\lambda$[B\_] $\lambda$\dd[B\dd\_], 
$\xi$[-a]$\sigma$[a, B, B\dd]] // TC
}
\end{flushleft}

Finally we eliminate the spinor indices in the
previous expression.
\begin{flushleft}
\Mathsize
\stepcounter{mathex}
\MathIn{mathex}{\\ d$\Xi$1 =ContractSolderingForm[\%, IndicesOf@Spin] // TC;}
\end{flushleft}

We study next the part containing the covariant derivatives of $\lambda_{A}$
(the expression {\tt d$\Xi2$}). 
This expression is a linear combination of spinors of the form $\nabla_{BC'}\lambda_A\nabla_{CB'}\bar\lambda_{A'}$ whose tensor counterpart has rank 3. 
This is the tensor we define next.

\begin{flushleft}
\Mathsize
\stepcounter{mathex}
\MathIn{mathex}{DefTensor[$\mathcal{Z}$[-a, -b, -c], M4]}
\DefTensorMessage{$\mathcal{Z}$[-a, -b, -c]}
\end{flushleft}

By definition

\begin{flushleft}
\Mathsize
\stepcounter{mathex}
\MathIn{mathex}{PutSolderingForm@$\mathcal{Z}$[-a, -b, -c] }
\MathOut{mathex}
{\tt $\mathcal{Z}_{\rm abc} \sigma^{\rm a}{}_{\rm AA'}
\sigma^{\rm b}{}_{\rm BB'} \sigma^{\rm c}{}_{\rm CC'}$}
\end{flushleft}

\begin{flushleft}
\Mathsize
\stepcounter{mathex}
\MathIn{mathex}{$\mathcal{Z}$rule =\\ 
 IndexRule[CDe[-B\_, -C\dd\_]@$\lambda$[-A\_] CDe[-C\_, -B\dd\_]@\
$\lambda$\dd[-A\dd\_], \%]}
\MathOut{mathex}
{\tt HoldPattern[$\nabla_{\underline{B}\underline{C'}}
\lambda_{\underline{A}}\nabla_{\underline{CB'}}\bar\lambda_{\underline{A'}}$]:{\tiny $\rightarrow$}\newline 
Module[\{a, b, c\}, 
$\mathcal{Z}_{\rm abc} \sigma^{\rm a}{}_{\rm AA'}
\sigma^{\rm b}{}_{\rm BB'} \sigma^{\rm c}{}_{\rm CC'}$]}
\end{flushleft}

We use now this rule in the expression for \icode{d$\Xi$2} getting

\begin{flushleft}
\Mathsize
\stepcounter{mathex}
\MathIn{mathex}{d$\Xi$2 = \newline 
ContractSolderingForm[d$\Xi$2 /. $\mathcal Z$rule, IndicesOf@Spin] // TC}
\MathOut{mathex} \\
{\tt  $\frac{1}{6}{\rm i} Gg_{\rm bdcf} \mathcal{Z}_{\rm a}{}^{\rm df} - 
\frac{1}{6}{\rm i}Gg_{\rm bfcd} \mathcal{Z}_{\rm a}{}^{\rm df} - \frac{1}{6}{\rm i} 
Gg_{\rm adcf}\mathcal{Z}_{\rm b}{}^{\rm df} + \frac{1}{6}{\rm i}Gg_{\rm afcd} \
\mathcal{Z}_{\rm b}{}^{\rm df} + \frac{1}{6}{\rm i}Gg_{\rm adbf} \
\mathcal{Z}_{\rm c}{}^{\rm df} - \frac{1}{6}{\rm i}Gg_{\rm afbd} 
\mathcal{Z}_{\rm c}{}^{\rm df}$}
\end{flushleft}

We combine now the values just found for \icode{d$\Xi$1} and \icode{d$\Xi$2}
and expand the tetra-metrics (see subsection \ref{subsec:tensor-spinor}). 
The final result is

\begin{flushleft}
\Mathsize
\stepcounter{mathex}
\MathIn{mathex}{d$\mathcal F$ == d$\Xi$1 + d$\Xi$2 /. TetraRule@g // TC;}
\end{flushleft}
 
The right hand side of this expression is a complicated tensor expression of 26 terms. 
It can be simplified though if we compute its double dual (we carry out the computation in two steps)

\begin{flushleft}
\Mathsize
\stepcounter{mathex}
\MathIn{mathex}{EqualTimes[\%, epsilong[-p, a, b, c]]}
\MathOut{mathex}
{\tt  $\eta_{\rm p}{}^{\rm abc} \triangledown_{\rm c}\mathcal{F}_{\rm ab} =  
\mathcal{Z}^{\rm a}{}_{\rm ap} - \mathcal{Z}^{\rm a}{}_{\rm pa} -
\frac{1}{2}S[\triangledown]_{\rm pa}\xi^{\rm a} + 3 \Lambda[\triangledown]\xi_{\rm p}$}
\end{flushleft}

\begin{flushleft}
\Mathsize
\stepcounter{mathex}
\MathIn{mathex}{EqualTimes[\%, epsilong[p, -d, -h, -f]/2]}
\MathOut{mathex} \\
{\tt  $\triangledown_{\rm d}\mathcal{F}_{\rm fh} -  
\triangledown_{\rm f}\mathcal{F}_{\rm dh} + \triangledown_{\rm h}\mathcal{F}_{\rm df} =
\frac{1}{2} \eta_{\rm dfhp} \mathcal{Z}^{\rm a}{}_{\rm a}{}^{\rm p} - 
\frac{1}{2}\eta_{\rm dfhp} \mathcal{Z}^{\rm ap}{}_{\rm a} + \frac{3}{2} \eta_{\rm dfha} 
\Lambda[\triangledown] \xi^{\rm a} - \frac{1}{4} \eta_{\rm dfhp} 
S[\triangledown]_{a}{}^{p} \xi^{a}$}
\end{flushleft}

\begin{flushleft}
\Mathsize
\stepcounter{mathex}
\MathIn{mathex}{\% // TFRicciToRicci // RicciToEinstein // Decomposition // TC}
\MathOut{mathex} \\
{\tt  $\triangledown_{\rm d}\mathcal{F}_{\rm fh} -  
\triangledown_{\rm f}\mathcal{F}_{\rm dh} + 
\triangledown_{\rm h}\mathcal{F}_{\rm df} =
\frac{1}{2} \eta_{\rm dfhp} \mathcal{Z}^{\rm a}{}_{\rm a}{}^{\rm p} - 
\frac{1}{2}\eta_{\rm dfhp} \mathcal{Z}^{\rm ap}{}_{\rm a} - 
\frac{1}{4}G[\triangledown]_{\rm a}{}^{\rm p} \eta_{\rm dfhp} \xi^{\rm a}$}
\end{flushleft}

This last equation coincides with (\ref{eq:sparling-equation}) and thus we conclude its validity. 
In addition from the spinor expression for  
$\mathcal{Z}_{abc}$ we easily deduce the algebraic property 
(\ref{eq:dominant-property}) if we express the null vectors $k^a_1$, $k^a_2$ and 
$k^a_3$ as tensor products of spinors of rank-1 
(see e.g. theorem 2.3.6 of \cite{STEWART}). \qed

The spinor form of (\ref{eq:sparling-equation}) has been used as the starting
point of a proof of the {\em positive mass theorem}. The rough idea is to prove that the integrals of the Einstein and Sparling 3-form over suitable hypersurfaces extending to 
{\em infinity} yield a positive quantity. This is straightforward for the Einstein 
3-form if the {\em dominant energy condition} on the matter is assumed, but it 
requires more efforts for the sparling 3-form. In fact one needs to make 
a special choice of the spinor $\lambda_A$ in order to ensure the positivity and
there is more than one way of achieving this (a good accounnt of the different
choices tried can be found in \cite{BERGQVIST-POSITIVITY}).    

\section*{Acknowledgements}
AGP is supported by the Research Centre of Mathematics of the University of
Minho (Portugal) through the ``Funda\c{c}\~ao para a Ci\^encia e a Tecnolog\'{\i}a (FCT) Pluriannual
Funding Program'' and through project CERN/FP/116377/2010. 
AGP also thanks the Erwin Schr\"odinger Institute in Vienna (Austria) 
where part of this work was carried out for hospitality and financial support 
under the program ``Dynamics of General Relativity''.
JMM was supported by the French ANR Grant BLAN07-1\_201699 entitled
``LISA Science'', and also in part by the Spanish MICINN projects
2008-06078-C03-03 and FIS2009-11893. Both authors thank Dr. Thomas B\"ackdahl 
for his tests of prior versions of the Spinors package and constructive criticism.








\bibliographystyle{elsarticle-num}
\bibliography{/home/alfonso/trabajos/BibDataBase/Bibliography}




\end{document}